\documentclass[%
reprint,
twocolumn,
groupedaddress,
amsmath,amssymb,
aps,
prb,
]{revtex4-1}

\usepackage{graphicx}% Include figure files
\usepackage{dcolumn}% Align table columns on decimal point
\usepackage{bm}% bold math
\usepackage{caption}
\usepackage{subcaption}
\usepackage{xcolor,soul}
\newcommand{\cri}{CrI$_3$}
\definecolor{red}{rgb}{1.0, 0.01, 0.24}
\definecolor{blue}{rgb}{0.0, 0.44, 1.0}
\begin{document}

%\preprint{APS/123-QED}

\title{Electronic correlation, magnetic structure and magnetotransport in few-layer \cri}

\author{Soumyajit Sarkar}
%\affiliation{Faculty for Physics, University of Duisburg-Essen, Lotharstr. 1, 47057 Duisburg, Germany}
\email{soumyajit.sarkar@uni-due.de}
\author{Peter Kratzer}
\email{peter.kratzer@uni-due.de}
\affiliation{Faculty of Physics, University of Duisburg-Essen, Lotharstr. 1, 47057 Duisburg, Germany}
\date{\today}
\begin{abstract}
Using density functional theory combined with a Hubbard model (DFT+{\textit U}), the electronic band structure of \cri~multilayers, both free-standing and enclosed between graphene contacts, is calculated. 
We show that the DFT+{\textit U} approach, together with the `around mean field' correction scheme, is able to describe the 
vertical magnetotransport in line with the experimental measurements of magnetoresistance in multi-layered \cri~enclosed between graphene contacts.
Moreover, by interpolating between different double-counting correction schemes, namely the `around mean field' correction and the fully localized limit, we show their importance 
%of an appropriate double-counting correction scheme within the DFT+$U$ formalism 
for describing both the band structure and the ground-state total energy consistently. 
Our description of the magnetic exchange interaction is compatible with the experimentally observed antiferromagnetic ground state in the bilayer \cri~and the transition to a ferromagnetic arrangement in a small external magnetic field.
Thus, using spin-polarized DFT+{\textit U} with an `around mean field' correction, a consistent overall picture is achieved.
\end{abstract}

\maketitle

%\tableofcontents
%%%%%%%%%%%%%%%%%%%%%%%%%%%%%%%%%%%%%%%%%%%%%%%%%%%%%%%%%%%%%%%%%%%%%%%%%%%%%%%%%%%%%%%%%%%%%%%%%%%%%%%%%%%%
%                  Introduction
%%%%%%%%%%%%%%%%%%%%%%%%%%%%%%%%%%%%%%%%%%%%%%%%%%%%%%%%%%%%%%%%%%%%%%%%%%%%%%%%%%%%%%%%%%%%%%%%%%%%%%%%%%%%
\section{Introduction}
Experimental realization of magnetism in two-dimensional (2D) atomic crystals has attracted the attention of many researchers \cite{Gibertini2019NatNanotech}, in particular because of its promising applications in spintronics. However, our basic understanding of these materials is still far from complete and poses many challenges to predictive electronic structure theory. 
Potential pitfalls and remedies will be exemplified by the present study of \cri{ }using density functional theory (DFT).  
Researchers were puzzled by the observation that \cri{ }is ferromagnetic (FM) as bulk material\cite{Tsubokawa1960JPhysSocJpn,Dillon1965}, while stacks of few atomic layers of \cri{ }show antiferromagnetic (AFM) interlayer exchange \cite{Huang2017Nature}. 
Electronic structure calculations with local or semi-local density functionals reproduced the ferromagnetism of bulk and single-layer \cri~\cite{Wang2011JCondens,Sivadas2019NanoLett,Lado20172DMater,Jiang2018NanoLett,Zheng2018Nanoscale}, but failed to describe the antiferromagnetic coupling in bilayers \cite{Jang2019PhysRevMaterials}. 
This discrepancy could be attributed to the structural anomalies \cite{Thiel2019Science,Ubrig2019} in thin films or to an insufficient description of their electronic structure by standard computational methods.
Obviously, knowledge of the exact energetic position of the Cr $3d$ orbitals and their contribution to the electronic band structure play a crucial role for resolving this issue.     
We show that an informed description of the electronic correlation at the Cr atoms 
using the DFT+{\textit U} approach not only reconciles the different magnetic properties
of bulk and thin layers of \cri, but is also mandatory for understanding magnetotransport effects observed in layered structures \cite{Klein2018Science,Wang2018NatureCommun}.  
By carefully selecting the appropriate theoretical tools and comparing their implications to available experimental data, we demonstrate that it is possible to 
assign a meaningful value to the Hubbard parameter {\textit U} and to
analyze electronic, magnetic and transport properties of this complex material in a common theoretical perspective.  

In this paper, we briefly review the methodology of DFT+\textit{U} calculations and discuss the band structure of bilayer CrI$_3$ for two alternative schemes of the double-counting correction. We will argue that the `around-mean-field' correction scheme gives results consistent with experiment, and corroborate this claim by presenting computational results for the magnetic structure of bi- layer and trilayer \cri. Finally, magnetotransport perpendicular to \cri{ }thin films between graphene contacts will be discussed on the basis of calculated band structures in conjunction with a model Hamiltonian.

%%%%%%%%%%%%%%%%%%%%%%%%%%%%%%%%%%%%%%%%%%%%%%%%%%%%%%%%%%%%%%%%%%%%%%%%%%%%%%%%%%%%%%%%%%%%%%%%%%%%%%%%%%%%
% Geometry and Method
%%%%%%%%%%%%%%%%%%%%%%%%%%%%%%%%%%%%%%%%%%%%%%%%%%%%%%%%%%%%%%%%%%%%%%%%%%%%%%%%%%%%%%%%%%%%%%%%%%%%%%%%%%%%
\section{Methodology}
As starting point of our theoretical investigation, we briefly summarize the basic knowledge of the atomic and electronic structure of \cri.
In its bulk form, \cri{ }is a layered compound with a high-temperature (HT) and a low-temperature (LT) phase that differ in the relative spatial arrangement of the stacked layer bundles (see section S2 in the supplementary material).
Each layer bundle consists of a central layer of Cr atoms forming a honeycomb lattice, sandwiched between two iodine layers. 
The Cr atom (valence configuration $4s^2, \, 3d^4$) is trivalent in \cri{ }and gives away three of its six valence electrons to the monovalent iodine atoms. 
The remaining three electrons are spin-polarized and occupy $t_{2g}$ orbitals of the majority spin channel, giving rise to a magnetic moment of $3 \mu_B$ per formula unit. The conduction band in the majority spin channel is formed by the unoccupied $e_g$ orbitals. In the minority spin channel, all five $3d$ orbitals of Cr are unoccupied and hybridize to form the conduction bands of minority spin. 
These qualitative features of the electronic structure of \cri{ }are already well reproduced in calculations with local or semi-local DFT functionals.
However, a quantitative description of the materials properties relies on knowledge of the precise energetic position of the Cr-derived bands. In particular, the size of the band gap in the majority spin channel has an important effect on the magnitude of the exchange interaction, as will be outlined below. 
Such an improved description can be achieved by the DFT+{\textit U} method \cite{Anisimov1991PhysRevB,Dudarev1998PhysRevB} with a suitably chosen value of the Hubbard parameter $U$ that describes the screened on-site Coulomb repulsion of the Cr $3d$ electrons. 
%PK here one could add more general remarks
This can be seen as part of a more general effort to enhance DFT calculations by a quantum many-body Hamiltonian that may contain the interatomic Hund exchange $J$ as well as the Stoner parameter $I$ as further parameters in addition to $U$. While the capabilities of model Hamiltonians have been explored e.g. in Ref.~\onlinecite{Ryee2018SciRep}, here we concentrate solely on the effect of $U$. 
Note that, even within a material of given chemical composition, the value of $U$ may vary for samples of different dimensionality, e.g. bulk or atomically thin films, due to variations of the electrostatic screening of the Coulomb interaction by the environment \cite{Roesner2016NanoLett}.
%However, 
Moreover, it is well known that calculations with the DFT+{\textit U} scheme require a double-counting (DC) correction, i.e., the Coulomb repulsion of $3d$ electrons already included in the DFT description must not be counted twice when adding the DFT+{\textit U} correction term, 
see e.g. Ref.~\onlinecite{Park2014PhysRevB} for an in-depth discussion.
Previous calculations ~\cite{Sivadas2019NanoLett,Lado20172DMater,Jiang2018NanoLett,Zheng2018Nanoscale,Jang2019PhysRevMaterials} of \cri~tacitly relied on the fully localized limit (FLL) for this DC correction, a scheme based on the consideration 
that each sublevel of the Cr $3d$ shell with magnetic quantum number $m$ is either fully occupied or empty. 
However, as we will discuss below, the description of the electronic structure by the FLL scheme may not be fully satisfactory, in particular when magnetotransport should be addressed.

%Yet, as we will show below, FLL leads to conclusions for the magnetic structure of few-layer \cri{ }that are incompatible with experimental observations.
Alternatively, we employ the `around mean field' (AMF) scheme \cite{Czyzyk1994PhysRevB} that 
becomes exact if the sublevels are all {\em equally} occupied with an average occupation $n_{\sigma}$ compatible with the spin density (per atom) suggested by the DFT treatment. 
Here, the exchange-correlation potential of DFT is considered as an effective mean field, thus motivating the name `around mean field' correction. 
A more detailed derivation of the two approaches can be found in section S1 in the supplementary material.
The shift of the Cr-derived energy bands introduced by the DC correction is reflected in the formulas\cite{Petukhov2003PhysRevB}
\begin{eqnarray}
 \Delta V_{\rm AMF}(m m' \sigma) &=& -(U-J) \left(\rho^{(\sigma)}_{m m'} - n_{\sigma} \delta_{m m'} \right) \label{eq:VDC-AMF} \\
 \Delta V_{\rm FLL}(m m' \sigma) &=& -(U-J) \left(\rho^{(\sigma)}_{m m'} - \frac{1}{2} \delta_{m m'} \right) \label{eq:VDC-FLL} 
\end{eqnarray}
The term $\rho^{(\sigma)}_{m m'}$ stands for the density matrix of each spin channel (indexed by $\sigma$) that is determined self-consistently during the DFT+{\textit U} calculation. 
The indices $m, m' \in [-2,2]$ refer to the magnetic quantum number within the 3d shell of Cr. 
In a diagonal representation, the diagonal elements of the density matrix, i.e. $m=m'$, have the meaning of occupation numbers. Consequently, the level shifts introduced by the $\Delta V$ may be positive or negative, depending on the average occupation of the levels. In the AMF scheme, the spin-dependent occupation $n_{\sigma}$ originating from the DFT calculation serves as reference point. 
We note that energy correction terms $\Delta E$ for both types of $\Delta V$ can be constructed (see Supplemental Material); and thus it is possible to calculate not only the band structure, but also the total energy in both approaches.

We perform spin-polarized DFT calculations using the full-potential all-electron code FHI-aims\cite{BLUM2009ComputPhysComm} with a highly accurate basis set constructed from atom-centered numerical orbitals. 
For the  exchange-correlation functional of DFT, we employ the generalized-gradient approximation (GGA) 
augmented with the Tkatchenko-Scheffler~\cite{Tkatchenko2009PhysRevLett} pairwise dispersive correction in order to account for the dispersive interaction.  
Spin-orbit interactions was taken into account where indicated, using the method of second-order variations, as described in Ref.~\onlinecite{Huhn2017PhysRevMaterials}. 
A $13 \times 13 \times 13$ and $11 \times 11 \times 1$ Monkhorst-Pack~\cite{Monkhorst1976PhysRevB,KrNe19} k-point mesh was used for Brillouin zone integration for the bulk and slab geometry, respectively. 
For the two-dimensional geometries a vacuum of length 20 \AA~was considered along the normal direction.  
The geometrical optimization calculations for each scheme, AMF or FLL,
were carried out until all the components of force on any atom is less than 5 meV/\AA. 
We employ the isotropic variant of DFT+{\textit U} \cite{Dudarev1998PhysRevB} treating $U_{\rm eff} = U - J$ as a single parameter. 
Treating the Hund exchange coupling $J$ as an independent parameters would offer an alternative approach to describe the electronic structure of CrI$_3$ that has been attempted by others \cite{Subhan2019} recently. 
In addition to varying $U_{\rm eff}$, we employ the linear interpolation scheme (see section S1 in the supplementary material) introduced by Petukhov {\textit et al.}\cite{Petukhov2003PhysRevB} with an additional parameter $\alpha \in [0,1]$ allowing them to switch smoothly between the limiting cases of AMF ($\alpha =0)$ and FLL $(\alpha = 1)$. 
Variations of the Cr--I bond length with the value of $U_{\rm eff}$ and with the double-counting correction scheme were found to be negligibly small, in accord with previous studies \cite{Wang2011JCondens}.

%%%%%%%  Figure %%%%%%
\begin{figure}[ht]
	\begin{subfigure}{0.45\linewidth}
		\centering
		\includegraphics[scale=0.35]{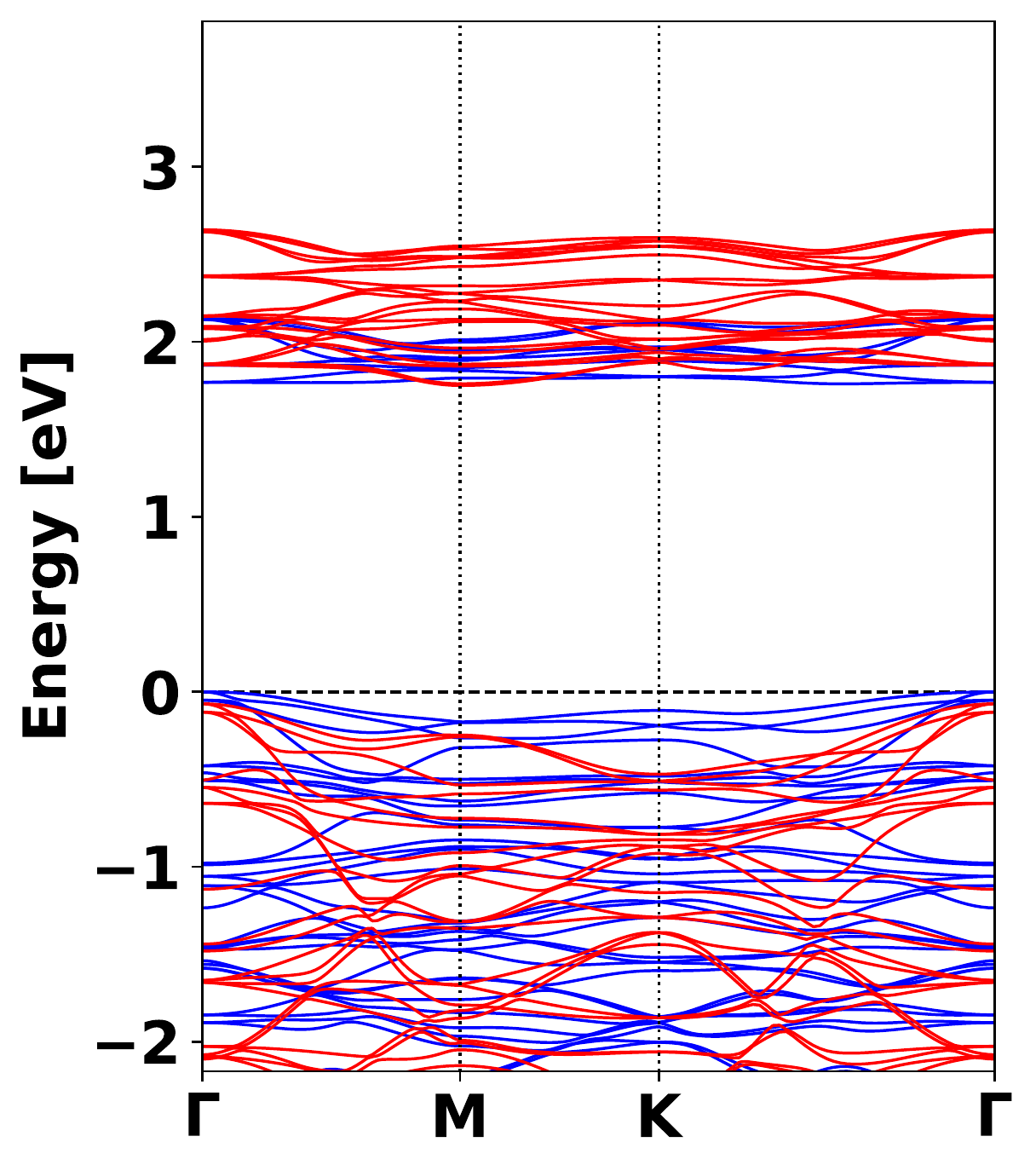}
		\caption{AMF correction, $\alpha=0$}
		\label{fig:bands_alpha0}
	\end{subfigure}
	\hfill
	\begin{subfigure}{0.45\linewidth}
		\centering
		\includegraphics[scale=0.35]{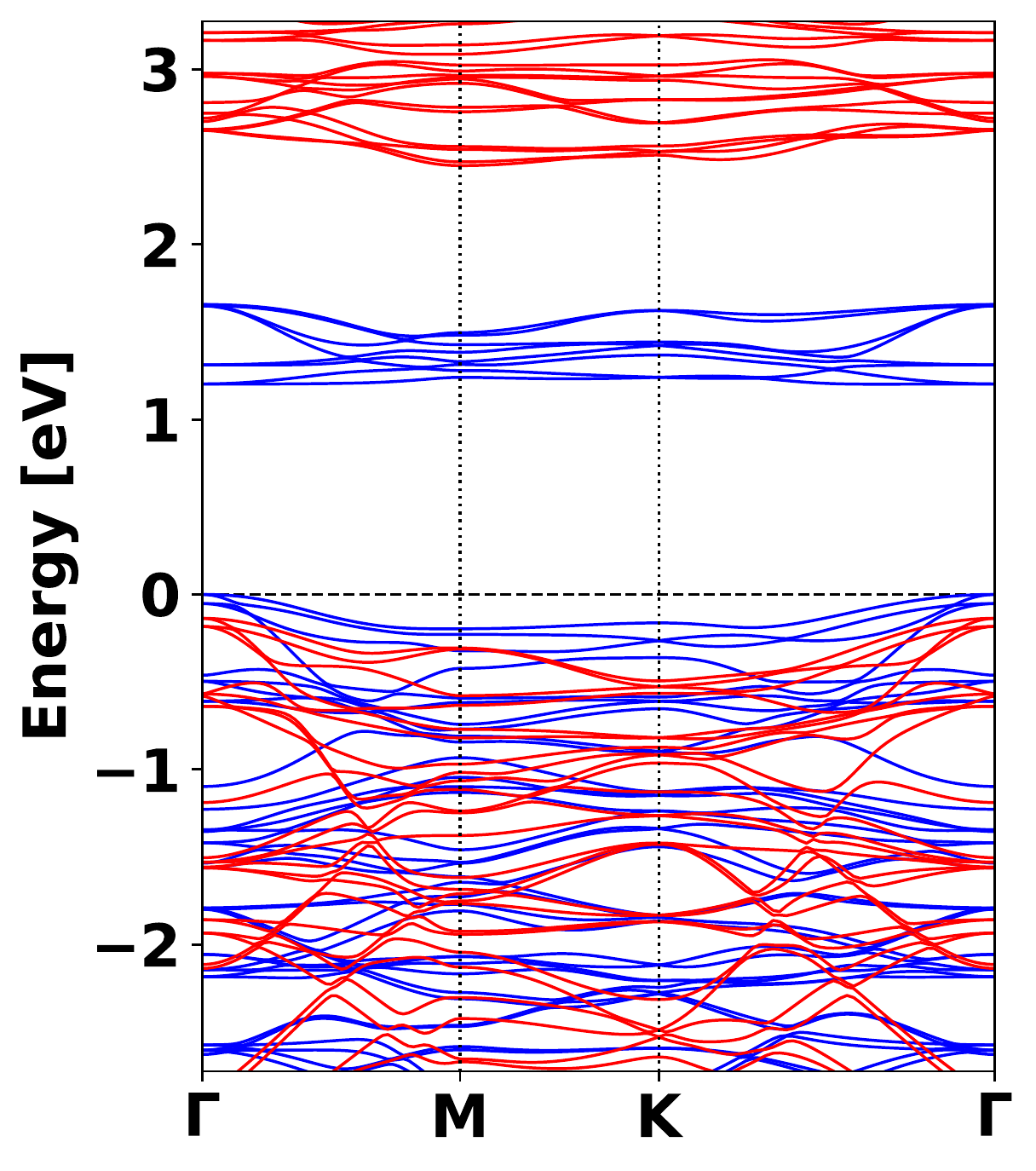}
		\caption{FLL correction, $\alpha =1$}
		\label{fig:bands_alpha1}
	\end{subfigure}

	\caption{Bands structure of a ferromagnetic bilayer of  CrI$_3$ (LT phase) for $U_{\rm eff}=1.7$ eV with around-mean-field double-counting correction, $\alpha=0$ (left) and in the fully localized limit, $\alpha=1$ (right). Blue and red lines indicate majority and minority spin bands, respectively. The lowest majority conduction bands belong to the $e_g$ representation and overlap with the minority spin conduction bands for $\alpha=0$, while they are energetically separated for $\alpha=1$. Zero of the energy axis is set at the top of the valence band.}
	\label{fig:bands_alpha_compared}
\end{figure}
%%%%%%%%%%%%%%%%%%

\section{Results}
{\em Electronic band structure} Typical results for the band structure in the two limiting cases of the AMF or FLL correction are shown in Fig.~\ref{fig:bands_alpha_compared}. 
Our particular interest is in the position of the conduction bands depending on the type of approximation used.
As a feature of the AMF scheme essential to the present study, we observe a large up-shift of the majority $e_g$ bands (Fig.~\ref{fig:bands_alpha_compared}a) compared to their position in pure DFT,   while the up-shift of the unoccupied minority bands is still moderate. 
As a consequence, both spin channels lie in an overlapping energy range. 
In contrast, the FLL scheme with any reasonably large $U_{\rm eff}$ parameter results in a {\em very large} up-shift of the minority bands, whereas the band gap in the majority spin channel remains small (Fig.~\ref{fig:bands_alpha_compared}b).
We will see below that the smaller gap in the FLL compared to the AMF scheme is closely related to preferred ferromagnetic ordering in the bilayer, whereas experiments rather point to a subtle energy balance between FM and AFM structures.
The origin and sign of the shifts can be understood from the expressions in Eq.s~(\ref{eq:VDC-AMF}) and (\ref{eq:VDC-FLL}):  
In the FLL scheme, the `neutral point' where shifts change sign is located at half-filling of a sublevel, as can be seen from the 1/2 in second term in the parentheses in Eq.~(\ref{eq:VDC-FLL}).
In the present case of \cri, it is mandatory to determine the `neural point' independently for each spin channel, thereby preserving the magnetic moment of 3$\mu_B$ per formula unit. 
With this prerequisite, the AMF correction turns out to be the better choice, as will be corroborated by the results presented below.

{\em Magnetic properties}
\cri\ is a magnetic semiconductor with a gap of 1.2 eV \cite{Dillon1965} in bulk 
and sizable intra-layer coupling of the Cr magnetic moments within one layer, but weaker couplings between adjacent layers. 
The latter may depend on stacking and thus differ in the LT and HT phase \cite{Jang2019PhysRevMaterials,Sivadas2019NanoLett}. 
The experimental observation that the magnetization in thin \cri{ }layers can be switched by an external magnetic \cite{Huang2017Nature,Thiel2019Science} or even an electric field \cite{Huang2018NatureNanotech,Jiang2018NatureMater} points to a subtle balance between AFM and FM couplings in this material.
For the intra-layer couplings, it has been suggested recently that there are two kinds of orbital-specific exchange interactions \cite{Kashin20202DMater}, an AFM interaction between occupied $t_{2g}$ orbitals at different Cr atoms, and an FM interaction between an occupied $t_{2g}$ orbital at one Cr and an unoccupied $e_g$ orbital at another Cr atom. 
We propose that this picture  extends to the interlayer interactions. 
Applying arguments from second-order perturbation theory, it has been shown that the magnitude of the $t_{2g} - e_g$ interactions is governed by the energetic distance $\Delta$ (see Eq.~7 in Ref.~\onlinecite{Kashin20202DMater}) between these two orbital groups, and thus on the energetic position of the $e_g$-type conduction bands in the majority spin. A high-lying $e_g$ band, as in Fig.\ref{fig:bands_alpha0} using AMF DC correction, therefore reduces the strength of the FM interaction, whereas this interaction gets enhanced in case of a low-lying $e_g$ conduction band, as in Fig.\ref{fig:bands_alpha1} using FLL. 
Thus it becomes clear that the experimentally observed AFM ground state of few-layer \cri{ }can be reproduced only by methods that correctly account for the energetic up-shift of the unoccupied majority spin levels. 

\begin{figure*}[ht]
	\begin{subfigure}{0.3\linewidth}
		\centering
		\includegraphics[scale=0.31]{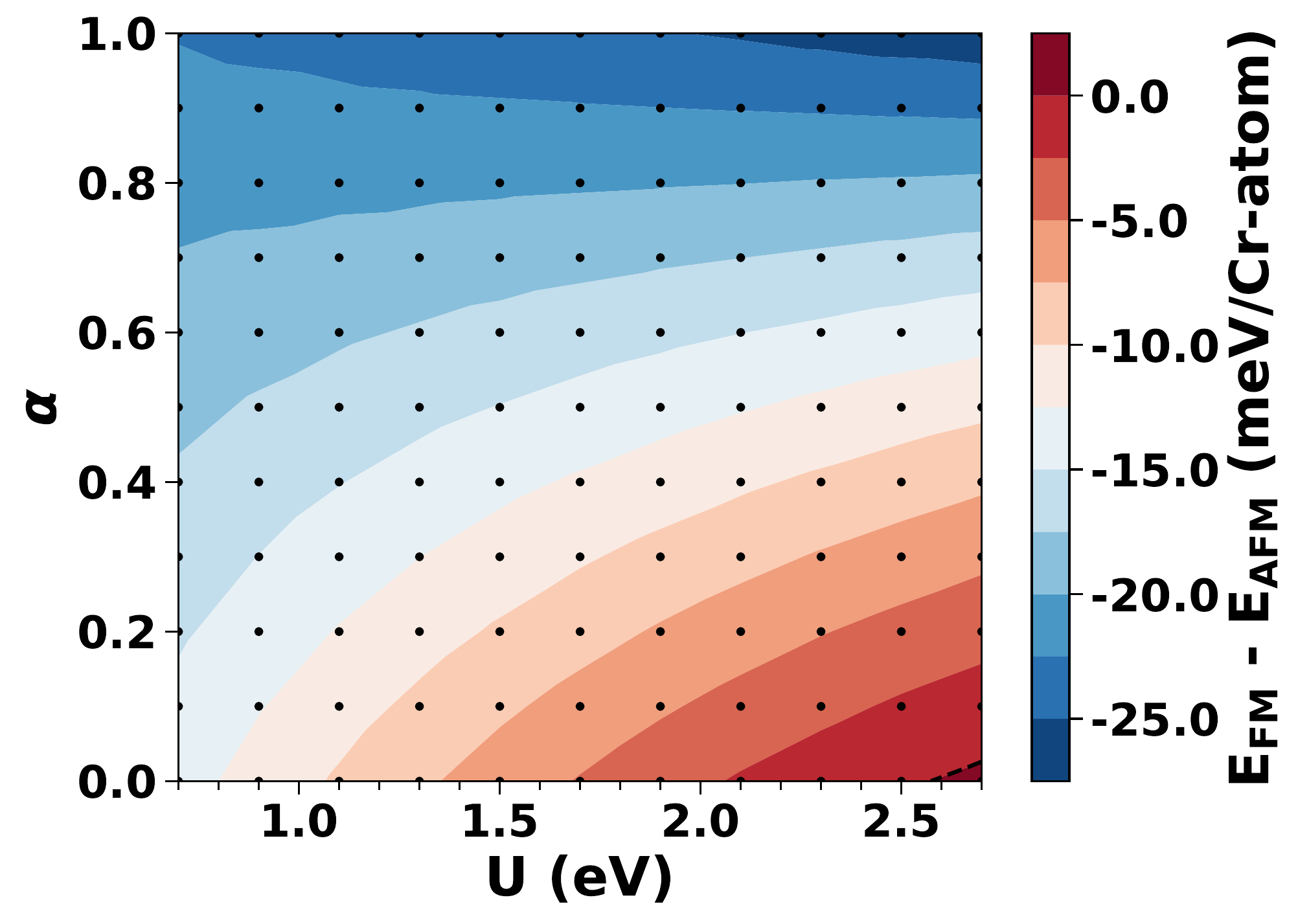}
		\caption{Bulk \cri{ } low-temperature phase}
		\label{fig:contour_bulk_LT}
	\end{subfigure}
	\hfill
	\begin{subfigure}{0.3\linewidth}
	\centering
			\includegraphics[scale=0.31]{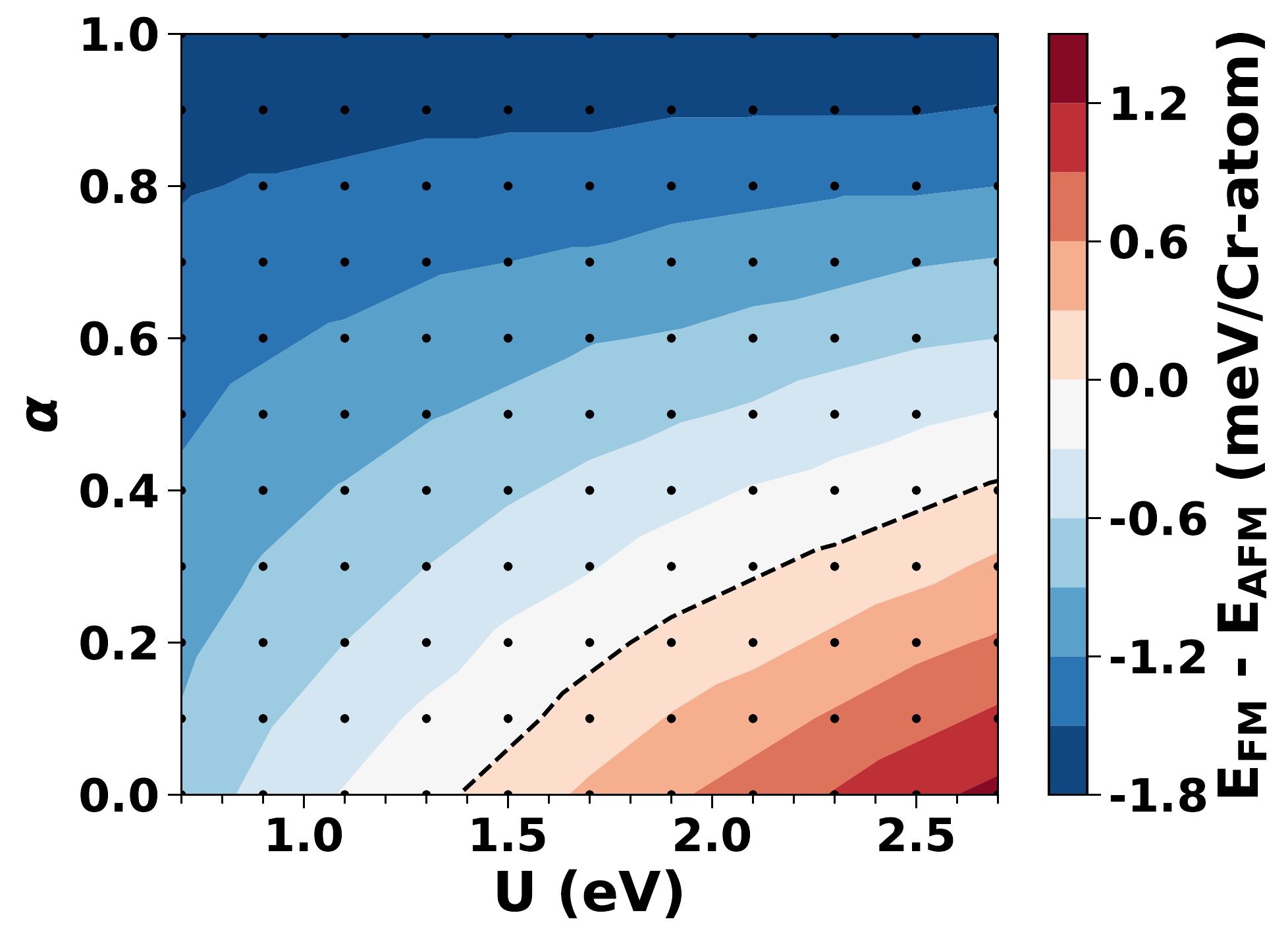}
	\caption{Bilayer \cri{ } low-temperature phase}
	\label{fig:contour_bilayer_LT}
	\end{subfigure}
	\hfill
	\begin{subfigure}{0.3\linewidth}
		\centering
		\includegraphics[scale=0.31]{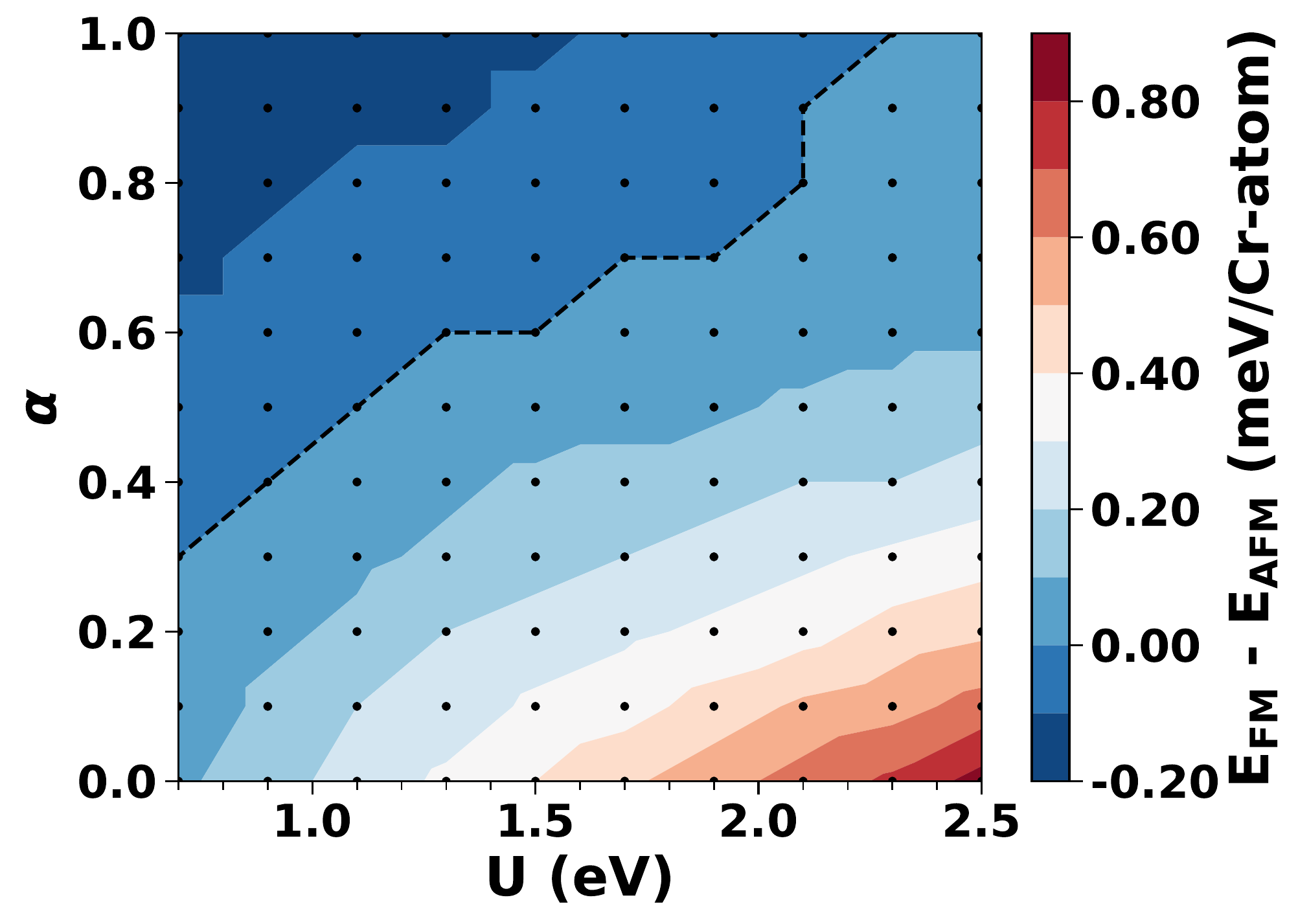}
		\caption{Bilayer \cri{ } high-temperature phase}
		\label{fig:contour_bilayer_HT}
	\end{subfigure}

	\caption{Contour plot of the total energy difference $\Delta E := E_{\rm FM} - E_{\rm AFM}$ between FM and AFM interlayer magnetic arrangements of low-temperature bulk (left), low-temperature bilayer (middle), and high-temperature bilayer (right) CrI$_3$ as function of the Hubbard $U_{\rm eff}$ and the DC correction parameter $\alpha$. Positive (Negative) values of the color bar indicate an AFM (FM) interlayer magnetic arrangement is favorable. The black-dashed line represents $\Delta E =0$~ contour line. Energy difference is expressed in meV/Cr-atom unit.}
	\label{fig:contour}
\end{figure*}

To provide evidence for this relation, we calculate the energy difference between FM and AFM interlayer magnetic arrangements as a function of two free parameters, $U_{\rm eff}$ and the interpolation parameter $\alpha$ controlling the type of DC correction. 
Fig.~\ref{fig:contour} shows contour plots of this energy difference on a 2D grid of parameter values, covering $U_{\rm eff}$ values in a physical reasonable range, between 0.7~eV and 2.5~eV.
This range is in agreement with previous studies: The constrained random-phase approximation method obtained 
$U_{\rm eff} = U - J = 2.9 - 0.7$eV for monolayer \cri{ } and $U_{\rm eff} = U - J = 2.0 - 0.7$eV \cite{Jang2019PhysRevMaterials}. 
A positive (negative) value of the energy difference $\Delta E := E_{\rm FM} - E_{\rm AFM}$ indicates an AFM (FM) interlayer magnetic arrangement has lower total energy. 
Next, we discuss and compare the results for bulk and for few-layer structures. 
In bulk \cri, FM interlayer magnetic interaction is found to be favored for almost all choices of $U_{\rm eff}$ and $\alpha$, see Fig.~\ref{fig:contour_bulk_LT}.
Only at high values of $U_{\rm eff}$ ($\geq 2.3$~eV), and close to the AMF limit, i.e., $\alpha \approx 0$, the energy difference approaches zero. 
Our results for the HT phase (not shown) of bulk \cri{ }are very similar to the LT phase: With the AMF correction scheme we predict FM magnetic arrangement to be stable independent of the value of  $U_{\rm eff}$ in the considered range.

%%%%%%%  Figure %%%%%%
%%%%%%%%%%%%%%%%%%

In the bilayer form of \cri, both in LT and HT phase, our calculations show a cross-over from FM to AFM magnetic arrangement as $U_{\rm eff}$ is increased. 
The LT structure of bulk \cri{ }would only become AFM at unrealistically high values of $U_{\rm eff} > 2.5$~eV. 
For LT bilayer and trilayer (shown in the Fig. S1 in the supplementary material) \cri, however, an AFM ground state is predicted for $U_{\rm eff}>1.4$~eV and small $\alpha$. 
In the AMF limit ($\alpha = 0$), we can describe the magnetic exchange energy per Cr atom $\Delta E$ by an empirical relationship $\Delta E = 1.07 \times 10^{-3} (U_{\rm eff} - 1.4 \mathrm{eV})$ for the bilayer and $\Delta E = 1.33 \times 10^{-3} (U_{\rm eff} - 1.4 \mathrm{eV})$ for the trilayer. 

Since $\Delta E$ is found to depend sensitively on $U_{\rm eff}$, experimental information about the magnetic switching behavior can be used to determine the physically meaningful value of  $U_{\rm eff}$ for a particular structure.
In the remainder of this paper, we will argue that GGA+{\textit U} with a value of $U_{\rm eff}=1.7$~eV, together with AMF DC correction, gives the best description of the electronic structure in few-layer structures that is in line with all experimental findings. 
The choice of $U_{\rm eff}=1.7$~eV for bilayers and trilayers is motivated by the experimentally observed magnetic field of 0.9~T required to induce the first transition of the magnetic structure when increasing the $B$-field\cite{Klein2018Science}. 
The Zeeman energy of 0.32~meV for flipping a magnetic moment of $\pm 3 \mu_B$ in this field equals the exchange energy $\Delta E$ obtained at $U_{\rm eff}=1.7$~eV in our calculation for the LT bilayer (Fig.~\ref{fig:contour_bilayer_LT}).
The second magnetic transition was experimentally observed at a field about twice as large, B=1.8~T\cite{Klein2018Science}. 
This is compatible with the idea that the AFM interlayer exchange interactions both to the layer above and below the central layer(s) must be overcome to induce switching in thicker samples.
Different interpretations of magnetic switching, for instance a relation with a possible structural change from the HT to the LT phase, as has been suggested by experimental work \cite{Thiel2019Science,Ubrig2019}, are also compatible with our calculations: The 
\cri{ }bilayer in the HT phase is predicted to be AFM in the AMF correction scheme at any reasonable $U_{\rm eff}$ value (cf. Fig.~\ref{fig:contour_bilayer_HT}), and could thus serve as starting point at $B=0$ for the magnetic transitions observed. 
If one assumed the HT structure, one would arrive at a slightly smaller value of $U_{\rm eff}=1.3$~eV at which the Zeeman energy of 0.32~meV matches the exchange energy $\Delta E$.
However, the HT phase is 1.38~meV per Cr atom higher in energy than the LT phase according to our calculations (with $U_{\rm eff} = 1.7$~eV and $\alpha = 0$), 
and we therefore prefer to continue the discussion under the assumption of the LT phase being present.
Elaborating further on the relation between dimensionality and $U_{\rm eff}$,
we note that a low value (e.g.  $U_{\rm eff}=1.3$eV as suggested in Ref. \onlinecite{Jang2019PhysRevMaterials}) may also be more appropriate inside bulk \cri{ }due to enhanced dielectric screening. 
This would not affect our claim that the GGA+{\textit U} scheme with AMF correction correctly captures the difference in magnetic properties of bulk and thin films, since this scheme
predicts the interlayer exchange interaction in bulk to be FM over a wide range of $U_{\rm eff}$ values both in the LT (Fig.~\ref{fig:contour_bulk_LT}) and the HT phase. 
 
 %%%%%%%  Figure %%%%%%
\begin{figure*}[ht]
	\begin{subfigure}{0.3\linewidth}
		\centering
		\includegraphics[scale=0.4]{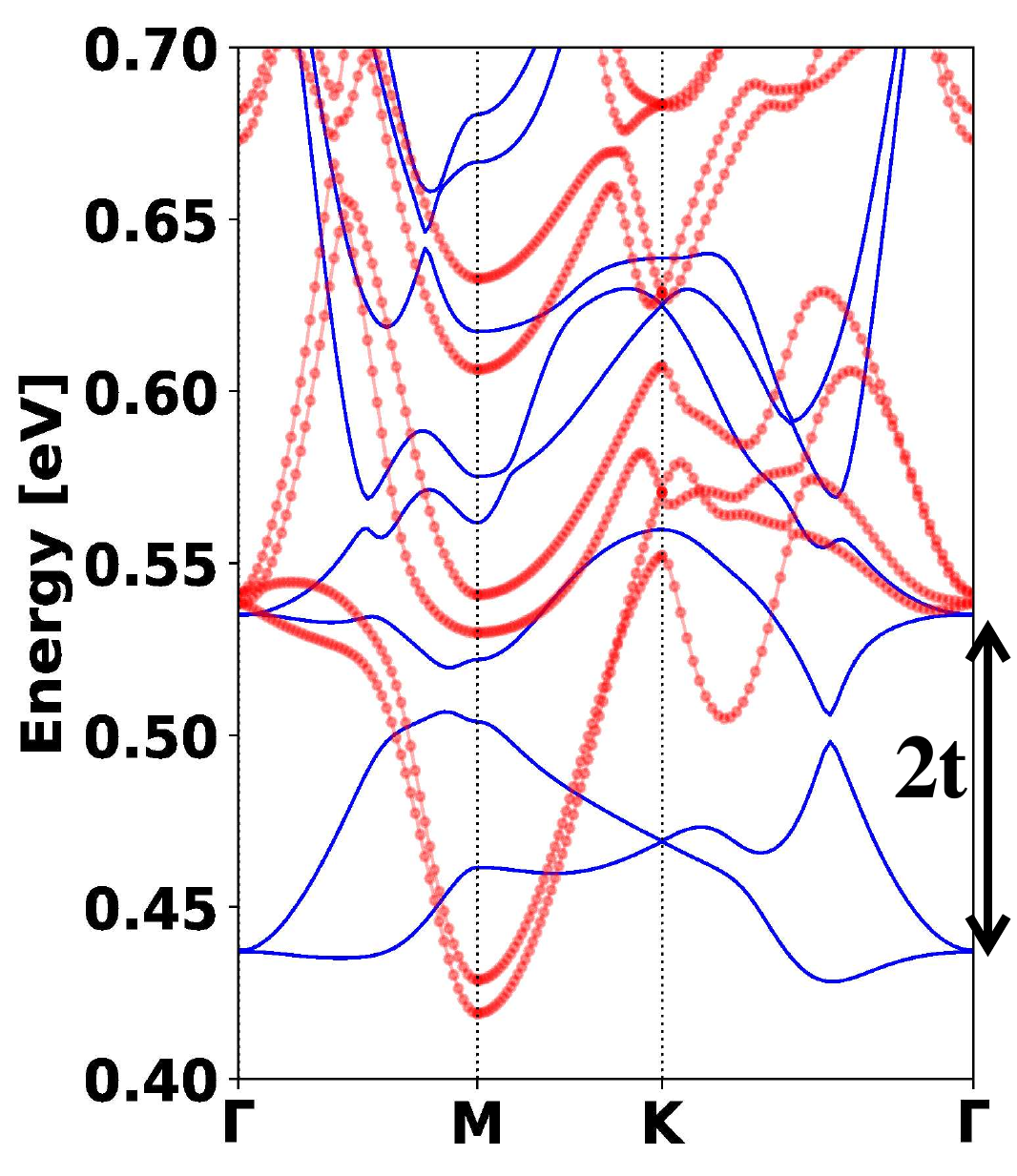}
		\caption{FM bilayer \cri}
		\label{fig:bilayer-bandstructureFM}
	\end{subfigure}
	\hfill
	\begin{subfigure}{0.3\linewidth}
		\centering
		\includegraphics[scale=0.4]{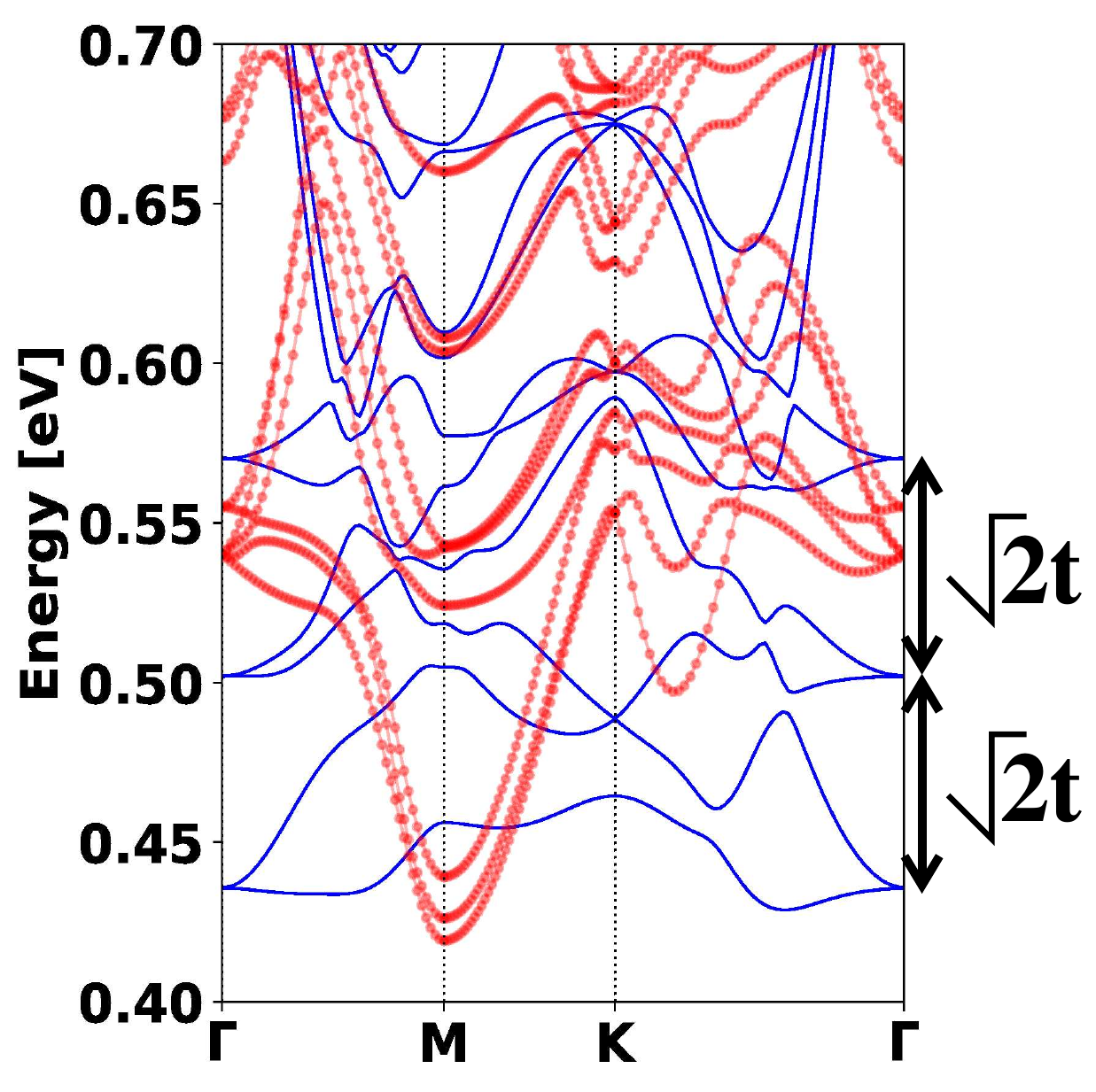}
		\caption{FM trilayer \cri}
		\label{fig:trilayer-bandstructureFM}
	\end{subfigure}
		\hfill
	\begin{subfigure}{0.3\linewidth}
		\centering
		\includegraphics[scale=0.4]{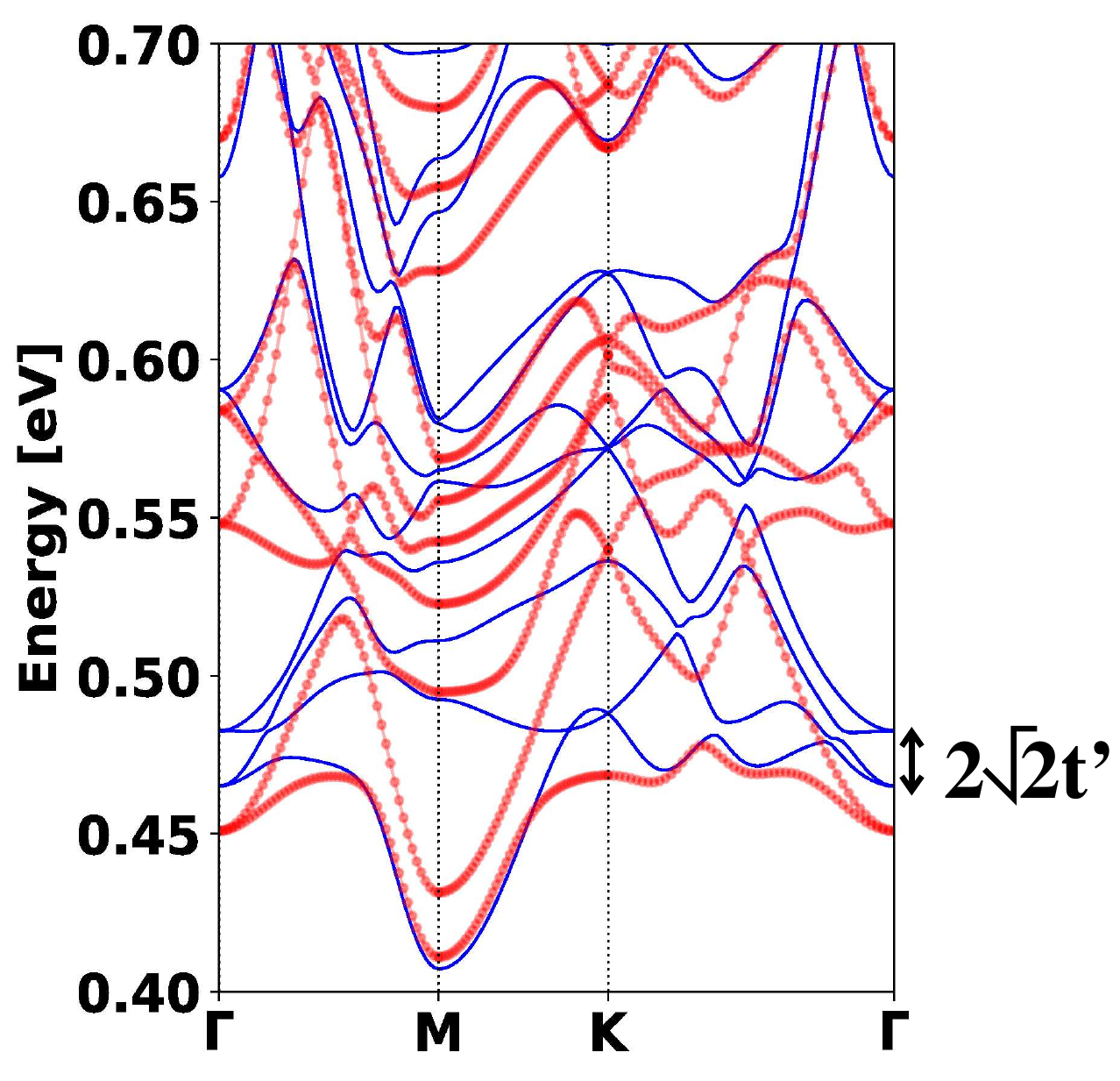}
		\caption{AFM trilayer \cri}
		\label{fig:trilayer-bandstructureAFM}
	\end{subfigure}

	\caption{Low-lying conduction bands of few-layer CrI$_3$ for a Hubbard $U_{\rm eff}=1.7$ eV and around-mean-field double-counting correction, $\alpha=0$. Band splittings used to parameterize a tight-binding model are indicated by the arrows.}
	\label{fig:bands-zoom}
\end{figure*}
%%%%%%%%%%%%%%%%%%

%%%%%%%%%%%%%%%%%%%%%%%%%%%%%%%%%%%%%%%%% 
{\em Magnetotransport}  
%%%%%%%%%%%%%%%%%%%%%%%%%%%%%%%%%%%%%%%%%
Tunneling magnetotransport perpendicular to the layers has been measured both in atomically thin samples \cite{Klein2018Science} and in samples several tens of nanometers thick \cite{Wang2018NatureCommun}, using graphene as contact material. In both studies, the tunneling current originated from electrons, rather than hole carriers. It was observed that the tunneling resistance is significantly lowered when the magnetization in the layers is aligned by an external magnetic field; this lowering amounts to a factor of 2, 4 and 6.5 in samples of 2, 3, and 4 layers (corresponding to magnetoresistance ratios of 100\%, 300\% and 550\%, see Ref.~\onlinecite{Klein2018Science}). Even larger magnetoresistance ratios, up to 8000\%, were observed in thicker samples\cite{Wang2018NatureCommun}. It is noteworthy that the magnetic fields at which the reduction of the resistance occurs are very similar, 0.9~T and 1.8~T in both types of samples, pointing to a common microscopic mechanism.
In the following, we will argue that these findings are compatible with majority and minority conduction bands at about the same energy, as found in our GGA+\textit{U} calculation with AMF correction, but cannot be reconciled with vastly different band energies, as typical for the FLL correction scheme.  

\begin{figure}[ht]
	\centering
	\includegraphics[scale=0.45]{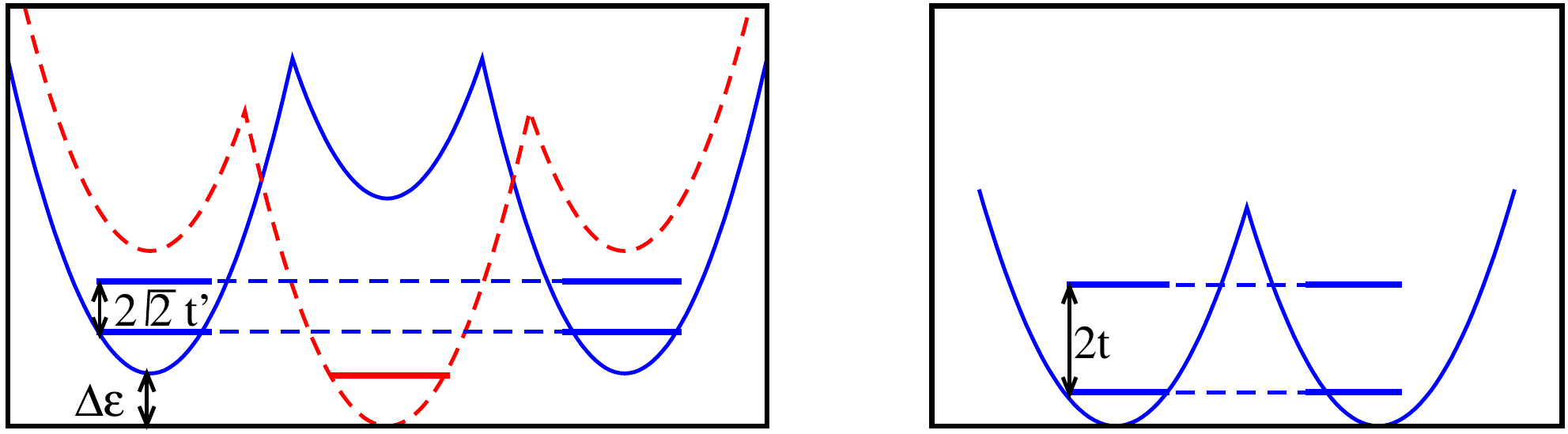}
	\caption{Schematic illustration of the exchange-correlation potential for spin-up (red) and spin-down (blue) conduction electrons in an AFM trilayer (left) and a FM bilayer (right) \cri. The splitting of energy levels due to tunneling (dashed horizontal line) between the layers is shown.}
	\label{fig:schematic}
\end{figure}

A qualitative explanation of the magnetoresistance effect can be given already from a schematic energy level diagram as illustrated in Fig.~\ref{fig:schematic}. 
The van der Waals gap between \cri~layers constitutes a barrier that must be overcome by the tunneling electrons. In an FM sample, the electrons will experience a periodic potential along the direction normal to the layers, with a period given by the layer thickness. The spin channel responsible for the conduction band minimum at the $\Gamma$ point formed by the orbitals with $e_g$ symmetry will contribute dominantly to the tunneling current. 
%, whereas electrons in the other spin channel make a negligibly small contribution.
This is different from the situation of antiparallel magnetization between adjacent layers: 
In the limit of the number of layers being large, electrons of either spin will contribute equally to the tunneling current, since spin majority electrons in one layer are minority electrons in the next layer, and vice versa; thus both electron spin states are equivalent. However, despite the two channels contributing, the overall current will be lower than in the ferromagnetic case because the tunneling barriers are higher.

%%%%%%%  Figure %%%%%%
%%%%%%%%%%%%%%%%%% 
For a more quantitative analysis, we have worked out a toy model in form of a tight-binding Hamiltonian (see section S4 in the supplementary material) with two hopping amplitudes $t$ and $t'$ for parallel and antiparallel layer magnetization, respectively.  
It is even possible to obtain estimates for these parameters from the GGA+\textit{U} band structures. 
Comparing level splittings obtained from the eigenvalues of the tight-binding Hamiltonian with the calculated band structures in Fig.~\ref{fig:bilayer-bandstructureFM} and \ref{fig:trilayer-bandstructureFM} yields $t \approx 50$~meV as best estimate. 
In a similar way, the value $t' \approx 6$~meV can be derived by comparing the analytically obtained eigenvalues with the calculated band structure in Fig.~\ref{fig:trilayer-bandstructureAFM}.

According to the Landauer-B{\"u}ttiker approach \cite{DattaSupriyo2005Qt:a} to transport, the conductance is $G \propto e^2/h \, T$ with the transmission probability $T =T^{\rm P}$ or $T =T^{\rm AP}$ for parallel or antiparallel magnetization of the layers, respectively. The latter two quantities can be expressed in terms of the Green function of the tunneling electrons, and is hence related to the hopping amplitudes $t$ and $t'$ in the Hamiltonian (see section S4 in the supplementary material for details). 
The transmission probability through a van-der-Waals gap between two layers with parallel spin alignment can be shown to be  $T_2^{\rm P} \propto |t|^2$.
For a ferromagnetic stack of $N$ layers, this expression generalizes to $T_N^{\rm P} \propto |t|^{2(N-1)}$. 
For a stack with alternating spin orientation, tunneling through two subsequent van der Waals gaps yield a probability $T_3^{\rm AP} \propto |t'|^4$. 
For $N$ layers ($N$ odd), the total probability is thus $T_N^{\rm AP} \propto |t'|^{2(N-1)}$. 
For a spin-filter model similar to the one considered in Ref.~\onlinecite{Klein2018Science} the magnetoresistance ratio is given as 
$$
\frac{G^{\rm high}}{G^{\rm low}} = \frac{T_N^{\rm P} }{T_N^{\rm AP} } \to \frac{|t|^{2(N-1)} }{2 |t'|^{2(N-1)}} = \frac{1}{2} \left| \frac{t}{t'} \right|^{2(N-1)}
$$ 
where the factor 2 stems from both spin channels contributing to transport in the antiparallel case. 
By inserting the values of $t=50$meV and $t'=6$meV from our band structure calculations, we obtain ratios of 35 and 2400 for bi- and trilayers, respectively. 
We note that a modest increase of the magnetoresistance ratio with the number of layers, as observed in the experiment~\cite{Klein2018Science}, can be obtained only if $t$ and $t'$ are of comparable magnitude, which in turn implies that the potentials experienced by spin-up and spin-down conduction electrons must be rather similar. 
For the AMF correction (see e.g. Fig.~\ref{fig:bands_alpha0}) we are close to achieving this condition by fine-tuning around $U_{\rm eff} \approx 1.7$eV, while this goal is out of reach in calculations with the FLL correction (Fig.~\ref{fig:bands_alpha1}), lending additional support to the AMF correction being the appropriate scheme for \cri. 

%%%%%%%  Figure %%%%%%
\begin{figure}[ht]
	\begin{subfigure}{0.45\linewidth}
		\centering
		\includegraphics[scale=0.35]{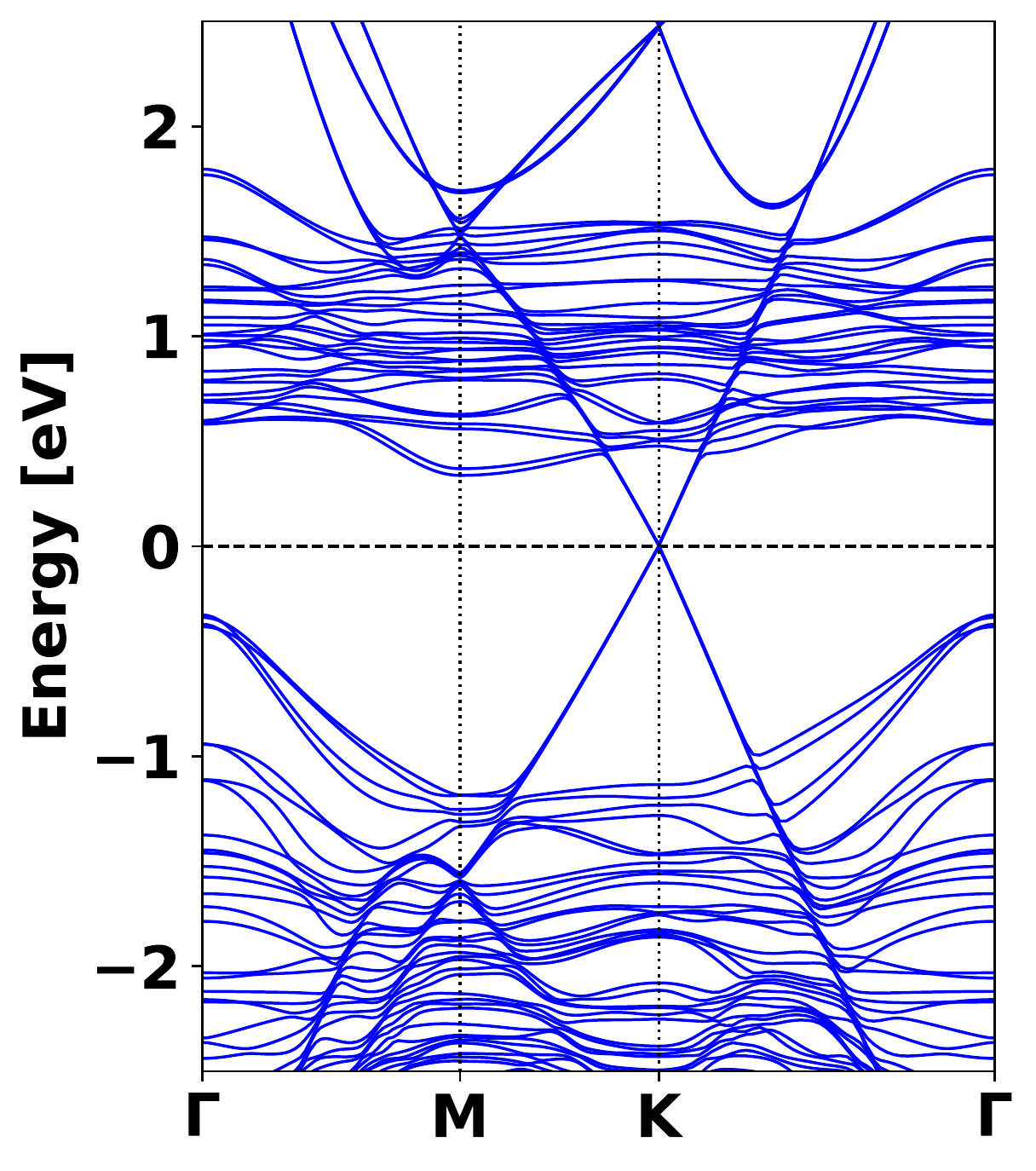}
		\caption{FM Bilayer \cri{ }with graphene contacts}
		%\label{fig:}
	\end{subfigure}
	\hfill
	\begin{subfigure}{0.45\linewidth}
		\centering
		\includegraphics[scale=0.35]{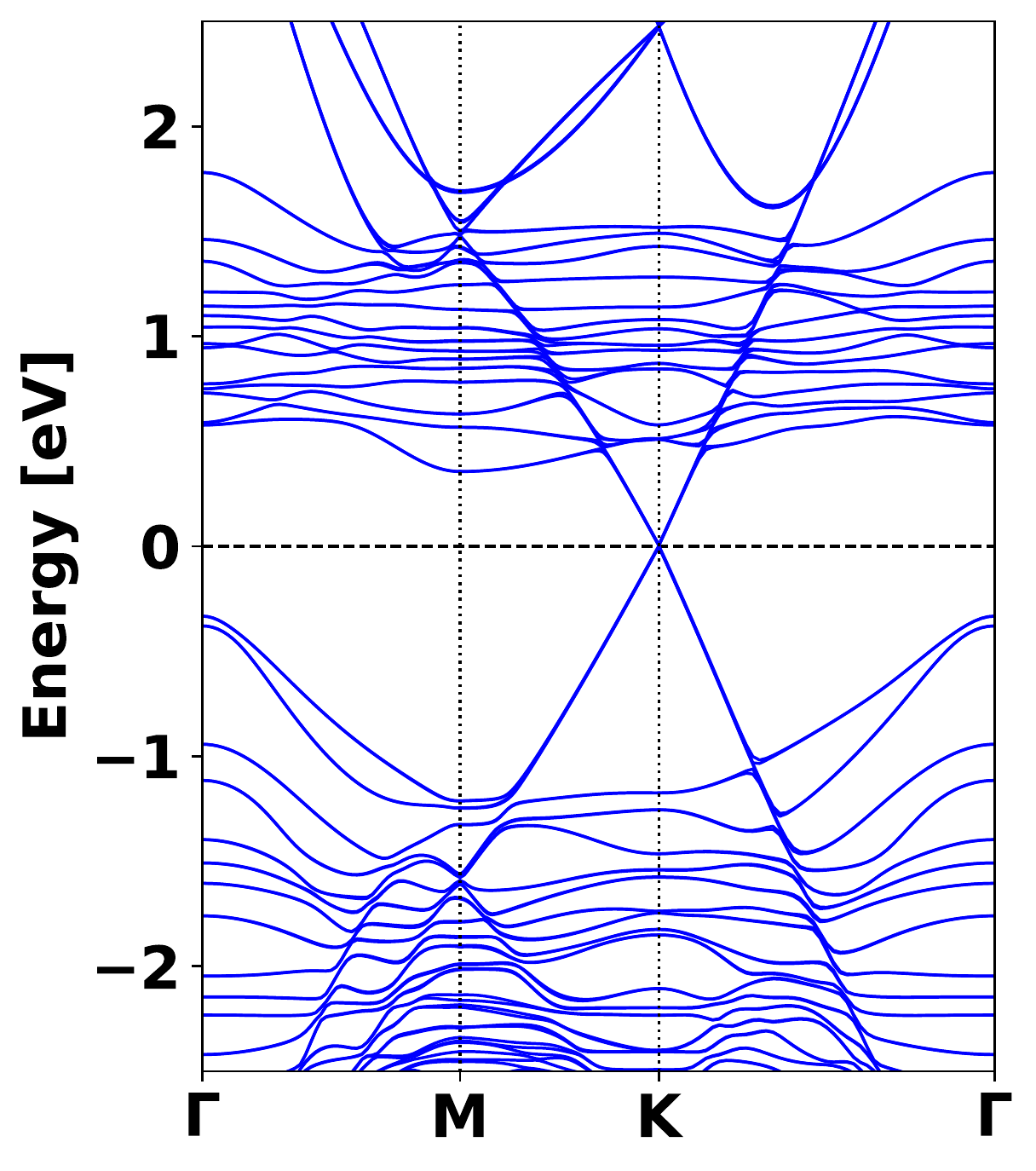}
		\caption{AFM Bilayer \cri{ }with graphene contacts}
		%\label{fig:}
	\end{subfigure}

	\caption{Band structure including spin-orbit interaction for a stack of bilayer CrI$_3$ between two graphene contacts for a Hubbard $U_{\rm eff}=1.7$ eV and around-mean-field double-counting correction, $\alpha=0$.}
	\label{fig:bands-graphen}
\end{figure}
%%%%%%%%%%%%%%%%%%

Eventually, we analyze the role of the graphene contacts. Clearly, the electrons need to tunnel from the graphene into the first layer of \cri, and likewise tunnel out of the terminating \cri{ }layer into the graphene at the other contact. 
Our GGA+{\textit U} calculations including graphene top and bottom contacts show that the structure of the valence and conduction bands in \cri{ }is essentially the same as in the free-standing bilayer. 
The weak van-der Waals interaction has a minor effect on the band structure, mostly mediated by the in-plane lattice constants which we assume to adjust to obtain a commensurate superstructure. 
The Dirac point of graphene, and thus the Fermi energy $\varepsilon_{\rm F}$ of (undoped) graphene, falls into the band gap of \cri.
Without applied voltage, the off-set between the conduction band edge and $\varepsilon_{\rm F}$ amounts to 0.42~eV at the M point and 0.44~eV at the $\Gamma$-point  (formed by $e_g$ orbitals of Cr). 
Under operating conditions, the potential drop between the graphene injector and the \cri{ } central layers will pull the \cri{ } conduction band minima closer to the Fermi energy in graphene. Our value for the off-set, being an upper bound for the injection barrier, is thus compatible with the activation energy of 0.15~eV found in temperature-dependent measurements of the magnetotransport  \cite{Wang2018NatureCommun}. 
   
When analyzing the magnetoresistance ratio of multilayer samples between graphene contacts, one should take into consideration that at least the outermost layer, but possibly even two or three  subsurface layers of \cri, display magnetic ordering different from bulk. This is supported by experiments \cite{Wang2018NatureCommun} that report switching of the magnetic structure by applied magnetic fields similar to those in the few-layer samples. 
From the perspective of our GGA+{\textit U} method with AMF correction, one is tempted to assume that 
some critical value of $U_{\rm eff}$ may exist for Cr atoms in a surface layer, in analogy to the transition for bilayer \cri~observed in Fig.~\ref{fig:contour_bilayer_LT}. If the critical $U_{\rm eff}$ value is exceeded, the interlayer coupling will turn AFM. 
It is plausible that Cr atoms in a surface layer should be described by a quite large $U_{\rm eff}$, in between the values for bulk and bilayer, since dielectric screening is reduced near a surface or interface. 
Since this effect is to be expected {\em both near the upper and lower} interface of the sample with the graphene contacts, it could also help to explain the very large values of the magnetoresistance ratio observed in the thicker samples \cite{Wang2018NatureCommun}. 

%%%%%%%%%%%%%%%%%%%%%%%%%%%%%%%%%%%%%%%%%%%%%%%%%%%%%%%%%%%%%%%%%%%%%%%%%%%%%%%%%%%%%%%
\section{Conclusions}
%%%%%%%%%%%%%%%%%%%%%%%%%%%%%%%%%%%%%%%%%%%%%%%%%%%%%%%%%%%%%%%%%%%%%%%%%%%%%%%%%%%%%%%
In conclusion, we have demonstrated that the GGA+\textit{U} approach to electronic structure, when applied in conjunction with double-counting corrections of `around mean field' type, gives a consistent overall picture of the electronic structure of \cri, its magnetic and transport properties.  
This approach carries over to other chromium trihalides; moreover, we believe that 
a fresh look at double counting correction schemes other than the commonly used `fully localized limit' is in place when extending the GGA+\textit{U} approach beyond transition metal oxides.
This finding may have important consequences for calculations aiming at the prediction of magnetic exchange parameters and of magnetic ordering temperatures. 

%%%%%%%%%%%%%%%%%%%%%%%%%%%%%%%%%%%%%%%%%%%%%
\section{Acknowledgments}
%%%%%%%%%%%%%%%%%%%%%%%%%%%%%%%%%%%%%%%%%%%%%
The authors gratefully acknowledge the funding of this project by computing time provided by the Paderborn Center for Parallel Computing (PC2).
\bibliography{Ref_merged}
\end{document}